\documentstyle[pra,aps]{revtex}

\begin{document}
\draft

\def\lcross{{\mathchar'26\mkern-9mu\lambda}}
\def\displayfrac#1#2{\frac{\displaystyle #1}{\displaystyle #2}}
\def\vcross{\vee\!\!\!\!\!\!-}

\def\bp{\mbox{\boldmath $p$}}
\def\br{\mbox{\boldmath $\rho$}}
\def\bz{\mbox{\boldmath $\zeta$}}

\title{Peres-Horodecki separability criterion for continuous variable
systems\thanks{This work was presented as part of an invited talk at the 6th
International Conference on {\em Squeezed States and Uncertainty Relations},
Naples, May 24 -- 29, 1999.}}
\author{R. Simon}
\address{The Institute of Mathematical Sciences, Tharamani,
Chennai 600 113, India}
\date{\today }
\maketitle
\begin{abstract}
The Peres-Horodecki criterion of positivity under partial transpose is studied
in the context of separability of bipartite continuous variable states. The
partial transpose operation admits, in the continuous case, a geometric
interpretation as mirror reflection in phase space. This recognition leads to
uncertainty principles, stronger than the traditional ones, to be obeyed by
all separable states. For all bipartite Gaussian states, the Peres-Horodecki
criterion turns out to be necessary and sufficient condition for separability. 
\end{abstract}


%\narrowtext
%\baselineskip19pt

Entanglement or inseparability is central to all branches of the emerging
field of quantum information and quantum computation\cite{bennett}. A
particularly elegant criterion for checking if a given state is separable or
not was proposed by Peres\cite{peres}. This condition is necessary and
sufficient for separability in the $2 \times 2$ and $2 \times 3$ dimensional
cases, but ceases to be so in higher dimensions as shown by
Horodecki\cite{horodecki}. 

While a major part of the effort in quantum information theory has been in
the context of systems with finite number of Hilbert space dimensions,
more specifically the qubits, recently there has been much interest in the
canonical continuous case\cite{c1,c2,c3,c4,c5,c6}. 
We may mention in particular the experimental
realization of quantum teleportation of coherent states\cite{uncond}. It
is therefore
important to be able to know if a 
given state of a bipartite canonical continuous 
system is entangled or separable.

With increasing Hilbert space dimension, any test for separability will
be expected to become more and more difficult to implement in practice.  
 In this paper we show that in the limit of infinite dimension, 
corresponding to continuous variable bipartite states,  the
Peres-Horodecki
criterion  leads to a test that is extremely easy to 
implement. 
  Central to our work is the recognition that  the partial transpose
operation  acquires, in the
continuous case, a beautiful geometric interpretation as 
{\em  mirror reflection in the Wigner phase space}.
 Separability forces  
on the  second moments (uncertainties) a restriction that is  
 stronger than the traditional uncertainty principle;  even commuting
variables need to  obey an uncertainty relation. This 
restriction is used to prove that the 
Peres-Horodecki criterion is  necessary and sufficient  
 separability condition for all bipartite Gaussian states.

Consider a single mode described by annihilation operator $\hat{a} = 
(\hat{q} + i\, \hat{p})/\sqrt{2}$,  
  obeying the standard commutation relation 
$[\hat{q}, \hat{p}] = i$, which is equivalent to $[\hat{a}, 
\hat{a}^{\dag}]=1$. 
There is a one-to-one correpondence between density operators and c-number
Wigner 
distribution functions $W(q, p)$\cite{wigner}.  
The latter are real functions over the phase space and satisfy  an
additional 
property coding the nonnegativity of the density operator.
 It follows  from the definition of Wigner distribution that the 
transpose operation $T$, 
 which takes every  $\hat{\rho}$ to its transpose $\hat{\rho}^{T}$,  
 is {\em equivalent to} a mirror reflection in
phase space:
\begin{equation}
\hat{\rho}\, 
\longrightarrow \,\hat{\rho}^{T} \;\Longleftrightarrow\; 
W(q, p)\, \longrightarrow \,W(q, -p).
\label{eq:eq1}
\end{equation}
Mirror reflection is not a canonical transformation in phase space, 
and cannot be implemented unitarily in the Hilbert space. This is 
consistent with the fact that while $T$ is linear at the density operator 
level, it is antilinear at the state vector or wave function  level.

Now consider a bipartite system of two modes described by annihilation
operators $\hat{a}_{1} = (\hat{q}_{1} + i\, \hat{p}_{1})/\sqrt{2}$ and
$\hat{a}_{2} = (\hat{q}_{2} + i\, \hat{p}_{2})/\sqrt{2}$. Let Alice be in 
possession of mode 1 and let mode 2 be in the possession of Bob. By 
 definition, a quantum state $\hat{\rho}$ of the bipartite system 
is separable if and only if $\hat{\rho}$ can be expressed in the form
\begin{equation}
\hat{\rho} = \sum_{j} p_{j} \,\hat{\rho}_{j1} \otimes \hat{\rho}_{j2},
\label{eq:eq2}
\end{equation}
with {\em nonnegative} $p_{j}$'s, where $\hat{\rho}_{j1}$'s and
$\hat{\rho}_{j2}$'s are density operators of the modes of Alice and Bob
respectively.  It is evident from (2) that partial transpose operation (i.e.,
transpose of the density matrix with respect to only the second Hilbert
space under Bob's possession), denoted $PT$, takes a separable density
operator {\em necessarily} into a nonnegative operator, i.e., into a
bonafide density matrix. This is the Peres-Horodecki separability
criterion.

In order to study the  partial transpose operation  in the 
 Wigner picture, it
is convenient to arrange the phase space variables and the hermitian
canonical operators into four-dimensional column vectors
\begin{eqnarray*}
\xi = 
\left( \begin{array}{cccc}
	q_{1}&
	p_{1} &
	q_{2}&
	p_{2}
	    \end{array}
	\right),\;\;\; \hat{\xi} = \left( \begin{array}{cccc}
	\hat{q}_{1}&
	\hat{p}_{1}&
	\hat{q}_{2}&
	\hat{p}_{2}
	    \end{array}
	\right).
\label{eq:eq3}
\end{eqnarray*}
The commutation relations take the compact form\cite{gaussian}
\begin{eqnarray}
[\hat{\xi}_{\alpha}, \hat{\xi}_{\beta}] & = & i\,\Omega_{\alpha \beta},
~~~ \alpha, \beta = 1, 2, 3, 4; \nonumber \\
\Omega & = & \left( \begin{array}{cc}
			J & 0\\
			0 & J
		    \end{array} \right),\;\;\; J = \left( \begin{array}{cc}
						0 & 1\\
						-1 & 0
						\end{array}\right).
\label{eq:eq4}
\end{eqnarray}

Wigner distribution and the density operator are related
through the definition\cite{wigner,gaussian}

\begin{eqnarray}
W(q,p) = \pi^{-2}\!\!\int\!\!d^{\,2}q' \langle q -
q^{\prime}|\,\hat\rho\, 
		    |q + q^{\prime}\rangle
\exp(2 i\, q^{\prime}\cdot p).
\end{eqnarray}

\noindent
where $q=(q_1,q_2),\; p=(p_1,p_2)$. 
It follows  from this definition that the
  partial transpose operation on the bipartite density
operator transcribes faithfully into the following transformation on the
Wigner distribution:
\begin{equation}
PT:\;\;\; W(q_{1}, p_{1}, q_{2}, p_{2})\; \longrightarrow \; W(q_{1}, p_{1}, 
q_{2}, -p_{2}).
\label{eq:eq5}
\end{equation}
This corresponds to a  mirror reflection which inverts the $p_{2}$
coordinate, leaving $q_{1}$, $p_{1}$, and $q_{2}$ unchanged:
\begin{eqnarray*}
PT:\;\;\; \xi \longrightarrow \Lambda \xi,\;\;\; \Lambda =\mbox{diag}(1, 1, 1, 
-1). \label{eq:eq6}
\end{eqnarray*}
And the Peres-Horodecki separability criterion reads: {\em if $\hat{\rho}$
is separable, then its Wigner distribution necessarily goes over into a
Wigner distribution under the  phase space mirror reflection $\Lambda$}.
  $W(\Lambda\xi)$, 
 like  $W(\xi)$, should possess the ``Wigner quality", for any separable
bipartite state.

The Peres-Horodecki criterion has important implications for the
uncertainties or second moments. 
 Given a bipartite density operator $\hat{\rho}$, let us define $\Delta 
\hat{\xi} = \hat{\xi} - \langle \hat{\xi} \rangle$, where $\langle 
\hat{\xi}_\alpha \rangle = \mbox{tr}\hat{\xi}_\alpha \hat{\rho}$.
The four components of $\Delta \hat{\xi}$
obey the same commutation relations as $\hat{\xi}$. Similarly, we define
$\Delta \xi_{\alpha} = \xi_{\alpha} - \langle \xi_{\alpha} \rangle$ where
 $\langle \xi_{\alpha} \rangle$ is average with respect to the Wigner
distribution $W(\xi)$, and it equals  $\langle \hat \xi
_\alpha\rangle$. The 
uncertainties are defined as 
the expectations of the hermitian operators $\{ \Delta\hat{\xi}_{\alpha}, 
\Delta\hat{\xi}_{\beta} \} =
(\Delta\hat{\xi}_{\alpha}\Delta\hat{\xi}_{\beta}+ 
\Delta\hat{\xi}_{\beta}\Delta\hat{\xi}_{\alpha})/2$:
\begin{eqnarray}
\langle 
\{ \Delta\hat{\xi}_{\alpha}, \Delta\hat{\xi}_{\beta} \} \rangle 
&=&  \mbox{tr}\left(\{ \Delta\hat{\xi}_{\alpha}, \Delta\hat{\xi}_{\beta} \} 
\hat{\rho}\right)\nonumber\\ 
& =& \int\!\! d^{\,4}\xi\, \Delta \xi_{\alpha}\, \Delta \xi_{\beta}\, W(\xi).
\label{eq:eq7}
\end{eqnarray}

Let us now arrange the uncertainties or variances into a $4 \times 4$ real
variance matrix $V$, defined through $V_{\alpha \beta} = \langle \{
\Delta\hat{\xi}_{\alpha}, \Delta\hat{\xi}_{\beta} \} \rangle$. Then we have
the following compact statement of the {\em uncertainty
principle}\cite{gaussian}:
\begin{equation}
V + \frac{i}{2}\,\Omega  \geq 0.
\label{eq:eq8}
\end{equation}
Note that (7) implies, in particular, that $V > 0$.

The uncertainty principle (7) is a direct
consequence of the 
commutation relation (3) and the nonnegativity of 
$\hat{\rho}$. It is equivalent to the statement that 
$\hat{Q}=\hat\eta \, \hat{\eta}^{\dagger}$, with 
$\hat\eta=c_1\hat{\xi_1}+c_2\hat{\xi_2}+c_3\hat{\xi_3}+c_4\hat{\xi_4}$, is 
nonnegative for every set of (complex valued) $c$-number coefficients 
$c_\alpha$, and hence $\langle \hat{Q} \rangle = \mbox{tr} 
(\hat{Q}\hat{\rho}) \geq 0$. Viewed somewhat differently, it is {\em 
equivalent} to the statement that for every pair of real four-vectors 
$d,d^{\,\prime}$ the hermitian operators
$\hat{X}(d)=d^{\,T}\hat{\xi}=d_1\hat{q_1}+ 
d_2\hat{p_1}+ d_3\hat{q_2}+ d_4\hat{p_2}$ and $\hat{X}(d^{\,\prime})
= d ^{\,\prime \,T}\hat{\xi}= 
d^{\,\prime}_1\hat{q_1}+ d^{\,\prime}_2\hat{p_1}+
 d^{\,\prime}_3\hat{q_2}+ d^{\,\prime}_4\hat{p_2}$ obey

\begin{eqnarray}
\langle ( \Delta\hat{X}(d) )^2\rangle &+& \langle ( 
\Delta\hat{X}(d^{\,\prime}) )^2\rangle \geq \left| d^{\,\prime 
\,T}\Omega \,d \right| \nonumber\\
&&\;\;=|d_1d^{\,\prime}_2-d_2d^{\,\prime}_1+d_3d^{\,\prime}_4
-d_4d^{\,\prime}_3|.
\end{eqnarray}
The right hand side equals $|\,[\hat X(d),\hat X(d^{\,\prime})]\,|$.
Under the Peres-Horodecki partial transpose the Wigner distribution 
undergoes mirror reflection, and it follows from (8) that the variances 
are changed to $V \to \tilde{V}=\Lambda V \Lambda$. Since $W(\Lambda\xi)$
has to be a 
Wigner distribution if the state under consideration is separable, we have

\begin{equation}
\tilde{V}+\frac{i}{2}\,\Omega \geq 0, \;\;\; \tilde{V} = \Lambda V \Lambda,
\end{equation}

\noindent
as a {\em necessary} condition for separability. We may write it also in the 
equivalent form

\begin{eqnarray}
V+\frac{i}{2}\,\tilde{\Omega} \geq 0, \;\;\; \tilde{\Omega} =
\Lambda 
\Omega \Lambda=\left( \begin{array}{cc} J & 0 \\ 0 & -J \end{array} \right),
\end{eqnarray}

\noindent
so that separability of $\hat{\rho}$ implies an additional restriction 
 that has the same form as (8), with 
 $| d^{\,\prime \,T}\Omega\, d |$ on the right hand side replaced  by 
$\left| d^{\,\prime \,T}\tilde{\Omega}\, d \right|$. Combined with (8), 
this restriction  reads

\begin{eqnarray}
\langle ( \Delta\hat{X}(d) )^2\rangle &+& \langle ( 
\Delta\hat{X}(d^{\,\prime}) )^2\rangle\nonumber\\ 
&&\!\!\!\!\!\!\!\!\!\!\!\!\!\!\!\!\geq 
|d_1d^{\,\prime}_2-d_2d^{\,\prime}_1| + 
|d_3d^{\,\prime}_4-d_4d^{\,\prime}_3|,\;\;\forall\,d,d^{\,\prime}.
\end{eqnarray}

\noindent
This restriction, to be obeyed by all separable states, is
generically  
stronger than the usual uncertainty principle (8).
 For instance, let  
 $\hat{X}(d)$ commute with $\hat{X}(d^{\,\prime})$, i.e.,  
 let  $d^{\,\prime\,T}\Omega \,d =0$. If the state  is
separable, then $\hat{X}(d)$ and 
$\hat{X}(d^{\,\prime})$ cannot both have arbitrarily small uncertainties
unless 
$d^{\,\prime \,T}\tilde{\Omega}\,d=0$ as well, i.e., unless
 $d_1d^{\,\prime}_2-d_2d^{\,\prime}_1= 0 
 =d_3d^{\,\prime}_4-d_4d^{\,\prime}_3$. As an example,   $\hat{X}
= \hat{x_1}+\hat{p_1}+\hat{x_2}+\hat{p_2}$ and $\hat{Y} = 
\hat{x_1}-\hat{p_1}- \hat{x_2}+\hat{p_2}$  commute, but
the sum of their uncertainties in any separable state is $\geq4$.

The Peres-Horodecki condition (11) can be simplified. 
Real linear canonical transformations of a two-mode system constitute the 
 ten-parameter real symplectic group $Sp(4,R)$. For 
every real $4 \times 4$ matrix $S \in Sp(4,R)$, the irreducible canonical 
hermitian operators $\hat\xi $ transform among themselves, leaving the
fundamental commutation relation (3) invariant:

\begin{eqnarray}
S \in Sp(4,R): \;\;\; S\Omega S^T &=& \Omega, \nonumber \\
\hat{\xi} \rightarrow \hat{\xi^{\,\prime}} 
&=& S\hat{\xi}, \;\;
\left[ \hat{\xi^{\,\prime}_\alpha},
\hat{\xi^{\,\prime}_\beta} \right] = i\,\Omega_{\alpha\beta}.
\end{eqnarray}

The symplectic group acts unitarily and irreducibly on the two-mode
Hilbert space\cite{gausson}.
Let $U(S)$ represent the (infinite dimensional) unitary  operator  
corresponding to $S \in Sp(4,R)$.
It transforms the bipartite state vector $|\psi\rangle$ to
$|\psi^{\,\prime}
\rangle=U(S)|\psi\rangle$, 
and hence the density operator $\hat{\rho}$ to $\hat{\rho^{\,\prime}}= U(S) 
\,\hat{\rho} \,U(S)^\dagger$. This transformation takes  a 
strikingly simple form in the Wigner description, and this is one 
reason for the effectiveness of the Wigner picture in handling  
canonical transformations:

\begin{equation}
S\!:\;\;\hat{\rho} \longrightarrow  U(S) \,\hat{\rho}\,U(S)^\dagger
\Longleftrightarrow W(\xi) \longrightarrow W(S^{-1}\xi).
\end{equation}

\noindent
The bipartite Wigner distribution simply transforms as a scalar field
under 
$Sp(4,R)$. 
It  follows from (6) that the variance matrix 
transforms in the following manner:

\begin{equation}
S \in Sp(4,R): \;\;\; V \rightarrow V^{\,\prime} = SVS^T.
\end{equation}

\noindent
The uncertainty relation (7) has an 
$Sp(4,R)$ invariant form (recall $S \Omega S^T = \Omega$). But 
  separable states have to respect not just (7), but also  
 the restriction (9), and this requirement
 is preserved  only under the six-parameter $Sp(2,R) 
\otimes Sp(2,R)$ subgroup of $Sp(4,R)$  corresponding to independent {\em 
local linear canonical transformations} on the subsystems of Alice and
Bob:

\begin{eqnarray}
S_{\mbox{\scriptsize{local}}} &\in& Sp(2,R) \otimes Sp(2,R) \subset
Sp(4,R): \nonumber
\\
S_{\mbox{\scriptsize{local}}} &=& \left( \begin{array}{cc} S_1 & 0 \\ 0 &
S_2 
\end{array} \right), \;\;\;\; S_1JS_1^T = J = S_2JS_2^T.
\end{eqnarray}

\noindent
It is desirable to 
cast the Peres-Horodecki condition (11) in an $Sp(2,R) \otimes Sp(2,R)$ 
invariant form. To this end, let us write the variance matrix $V$ in the 
block form

\begin{eqnarray}
V=\left( \begin{array}{cc} A & C \\ C^T & B \end{array} \right).
\end{eqnarray}

\noindent
The physical condition (7) implies $A\ge 1/4, \; B\ge 1/4$. 
 As can be seen from (14), the local group changes the blocks of $V$  
in the following manner:

\begin{eqnarray*}
A \rightarrow S_1AS^T_1, \;\;\;B \rightarrow S_2BS^T_2, \;\;\;C 
\rightarrow S_1CS^T_2.
\end{eqnarray*}
Thus, the $Sp(2,R) \otimes Sp(2,R)$ invariants associated with $V$ are
$I_1
=\mbox{det}\, A, \; I_2 = \mbox{det}\, B, \; I_3 = \mbox{det}\, C$ and
$I_4 =
\mbox{tr}\, A J C J B J C^{T}J\;$ (det$\,V$ is an obvious
invariant, but it is  a function of the $I_k$'s, namely  
$\mbox{det}\, V \,=\, I_1 I_2 + I_3\,^2 - I_4$). 

We claim that the uncertainity
principle (7) is equivalent to the   $Sp(2,R) \otimes Sp(2,R)$
invariant statement
\begin{eqnarray}
\mbox{det}\, A \; \mbox{det}\, B + \left (\frac{1}{4} - \mbox{det}\,
C\right )^2 &-&\,\,
\mbox{tr} (A J C J B J C^{T} J) \nonumber \\
       \ge &\frac{1}{4}& (\mbox{det}\, A \,+\, \mbox{det}\, B)\,.
\end{eqnarray}

\noindent
To prove this result, first note that (7) and (17) are equivalent for
variance matrices of the special form

\begin{eqnarray}
V_0=\left( \begin{array}{cccc}
a & 0 & c_1 & 0 \\
0 & a & 0 &c_2 \\
c_1 & 0 & b & 0 \\
0 & c_2 & 0 & b \end{array} \right).
\end{eqnarray}
But any variance matrix can be brought to this special form by effecting
a suitable local canonical transformation corresponding to some element of  
$Sp(2,R) \times Sp(2,R)$. 
In veiw of the manifest $Sp(2,R) \otimes Sp(2,R)$ invariant structure of
(17), it follows  that (7) and (17) are indeed equivalent for all variance
matrices.

Under the Peres-Horodecki partial transpose or mirror reflection, we have  
$V \rightarrow \tilde{V} = \Lambda V \Lambda$. That is, $C \rightarrow C
\sigma_3$ and $B \rightarrow \sigma_3 B \sigma_3$, while $A$ remains
unchanged  [$\sigma _3$ is the diagonal Pauli matrix: $\sigma_3 =
\mbox{diag}(1,-1)$]. As a consequence, $I_3=
\mbox{det}\, C$ flips  signature while
$I_1,I_2$ and $I_4$ remain unchanged. Thus, condition (9) for
$\tilde{V}$ takes a form identical to (17) with only the signature in
front of
det$\,C$ in the second term on the left hand side reversed. Thus the requirement 
that the variance matrix of a separable state has to obey  (9), in
addition to 
the fundamental uncertainty principle  (7), takes the  form 

\begin{eqnarray}
\mbox{det}\, A \; \mbox{det}\, B +  \left({\frac{1}{4}} - |\mbox{det}\,
C| \right)^2 &-&\,\,
\mbox{tr} (A J C J B J C^{T} J) \nonumber \\
       \ge &\frac{1}{4}& (\mbox{det}\, A \,+\, \mbox{det}\, B).
\end{eqnarray}

\noindent
{\em This is the final  form of our  necessary 
condition on the variance matrix of a 
separable bipartite state. 
This condition is  invariant not only under 
$Sp(2,R) \otimes Sp(2,R)$, but also under mirror reflection, as it should
be!  It constitutes a complete description of the
implication the Peres-Horodecki criterion has for the second moments.}

To summarise,  conditions (7), (8), and (17) are equivalent statements
of the fundamental
uncertainty principle, and hence will be satisfied by every physical
state. 
The mutually equivalent statements (9), (11), and (19) constitute  the
Peres-Horodecki criterion at the level of the second moments, 
and should necessarily be 
satisfied by every separable state. Interestingly, 
states with $\det C \ge 0$  definitely satisfy (19), which in this case is
subsumed by the physical condition  (17).  

For the standard form $V_0$, our condition (19)
reads
\begin{eqnarray*}
4(ab - c_1^{\,2})(ab - c_2^{\,2}) \ge  (a^2 + b^2)
+ 2|c_1 c_2| -  1/4.
\end{eqnarray*}

\noindent
But the point is that the separability check (19) can be applied
directly on $V$, with no need to go to the form $V_0$. 

We will now apply these results to Gaussian states.
 The mean values $\langle \hat{\xi}_\alpha\rangle$ can be changed at will
using local
unitary displacement operators, and so assume without loss of generality 
 $\langle\hat{\xi}_\alpha \rangle  = 0$.
A (zero-mean) Gaussian states is fully characterized by its second
moments, as seen from the nature of the  Wigner distribution
\begin{eqnarray*}
W(\xi) = \left( 4 \pi ^2 \,\sqrt{\det V}\right)^{-1} 
\exp\left(-\frac{1}{2}\xi^{\,T}V^{-1}\xi \right).
\end{eqnarray*}

\noindent
{\bf Theorem:} {\em The Peres-Horodecki criterion (19) is necessary and
sufficient condition for separability, for all bipartite Gaussian
states.}\\

We begin by noting, in view of the P-representation 
\begin{eqnarray*}
\hat\rho = \int\!\!d^{\,2}z_1 d^{\,2} z_2 P(z_1,z_2) |z_1\rangle \langle z_1|
					\otimes |z_2 \rangle \langle z_2|,
\end{eqnarray*}
that a state which is classical in the quantum optics sense (nonnegative 
$P(z_1,z_2)\,)$ is separable. Since the local group 
 $Sp(2,R) \otimes Sp(2,R)$ does not affect separability, any 
$Sp(2,R) \otimes Sp(2,R)$ transform of a classical state is separable too.  
Finally, a Gaussian state is classical if and only if $V - \frac{1}{2} \ge 0$.
 We will first prove  a pretty little result.\\

\noindent
{\bf Lemma:} {\em Gaussian states with  $\det C \ge 0$ are separable}.\\

\noindent
First consider the case $\det C > 0$.  We can arrange $a \ge
b,\;\; c_1\ge c_2 > 0$
in the special form $V_0$ in (18).  Let us do a local  canonical
transformation 
 $S_{\mbox{\scriptsize{local}}} = \mbox{diag} \,(x,x^{-1},x^{-1},x)$, 
corresponding to reciprocal local scalings (squeezings) at the Alice and
Bob ends, 
 and follow it by  $S_{\mbox{\scriptsize{local}}}^{\,\prime} =
\mbox{diag}\, 
(y,y^{-1},y,y^{-1})$, corresponding to common local scalings at these
ends. 
 We have

\begin{eqnarray*}
V_0 \to V_0^{\,\prime}=\left( \begin{array}{cccc}
y^2x^2a & 0 & y^2c_1 & 0 \\
0 & y^{-2}x^{-2}a & 0 &y{-2}c_2 \\
y^2c_1 & 0 & y^2x^{-2}b & 0 \\
0 & y^{-2}c_2 & 0 & y^{-2}x^{2}b \end{array} \right).
\end{eqnarray*}

\noindent
Choose $x$ such that $c_1/(x^2a - x^{-2}b) = c_2/(x^{-2}a - x^2b)$. That
is,
$x=[(c_1a + c_2b)/(c_2a + c_1b)]^{1/4}$. With this choice,  
  $V_0^{\,\prime}$ acquires such
a structure that it can be diagonalized by rotation through {\em equal}
amounts in the $q_1,q_2$ and $p_1,p_2$ planes:

\begin{eqnarray*}
V_0^{\,\prime} &\to& V_0^{\,\prime\prime} = \mbox{diag}\, (\kappa
_+,\,\kappa_+^{\,\prime},\,\kappa_-,\,\kappa_-^{\,\prime})\,;\\
\kappa_\pm &=& \frac{1}{2}y^2\left\{ x^{2}a+x^{-2}b \pm 
[(x^{2}a - x^{-2}b)^2 + 4 c_1^{\,2}]^{1/2}\right\}, \\
\kappa_\pm^{\,\prime} &=& \frac{1}{2}y^{-2}\left\{ x^{-2}a+x^{2}b \pm 
[(x^{-2}a - x^{2}b)^2 + 4 c_2^{\,2}]^{1/2}\right\}.
\end{eqnarray*}

\noindent
Such an equal  rotation is a canonical transformation;  it 
preserves the uncertainty principle, since it is canonical,  and the
pointwise
nonnegativity of the P-distribution, since it is a rotation. For our
diagonal
$V_0^{\,''}$, 
 the uncertainty
principle $V_0^{\,''} + \frac{i}{2} \Omega \ge 0$ simply reads that the 
product 
 $\kappa_-\kappa_-^{\,'} \ge 1/4$. It follows that we can choose $y$ such
that   $\kappa_-,\,\kappa_-^{\,'} \ge 1/2$ (for instance,
choose $y$ such
that $\kappa_-\, =\, \kappa_-^{\,'}$), i.e., $V_0^{\,''} \ge 1/2$. Since
$V_0^{\,'}$ and $V_0^{\,''}$ are rotationally related, this implies
$V_0^{\,'} \ge1/2$, and hence $V_0^{\,'}$ corresponds to positive
P-distribution or separable state. This in turn implies that the original
$V$ corresponds to a separable state, since $V$ and $V_0^{\,'}$ are
related
by local transformation. This completes proof for the case $\det C > 0$. 

Now suppose $\det C =0$, so that in $V_0$ we have $c_1\ge 0 = c_2$. Carry
out a local scaling corresponding to $S_{\mbox{\scriptsize{local}}} =
\mbox{diag}\, (\,
\sqrt{2a},\, 1/\sqrt{2a},\, \sqrt{2b},\, 1/\sqrt{2b}\,)$,
 taking $V_0 \to V_0^{\,'}$; the diagonal
entries of $V_0^{\,'}$ are $(2a^2,\, 1/2,\,2 b^2,\, 1/2)$, and the two 
nonzero off diagonal entries equal $2abc_1$.  With this form for
$V_0^{\,'}$, the  uncertainty
principle $V_0^{\,'} + \frac{i}{2} \Omega \ge 0$ implies $V_0^{\,'} \ge
1/2$, establishing separability of the Gaussian state. This completes 
proof of our lemma.

Proof of the main theorem is  completed as follows. We 
consider in turn the two distinct cases $\det C <0$ and $\det C \ge 0$.
 Suppose $\det C <0$. 
 Then there are two possibilities. If (19) is violated, then the
Gaussian state is definitely entangled since (19) is a necessary condition
for separability. If (19) is respected, then the mirror reflected state is
a physical Gaussian state with $\det C > 0$ (recall that mirror reflection
 flips the signature of $\det C$), and  is separable by the above lemma. 
This implies separability of the original state, since a mirror reflected
separable state is separable.  
Finally,  suppose $\det C \ge 0$.  Condition (19) is definitely 
satisfied since it is subsumed by the 
uncertainty principle (17) in this case. By our lemma,  the state is
separable. This completes
proof of the theorem.

We have  worked in the Wigner picture. But,
 the geometric interpretation of the  partial transpose as mirror
reflection in phase  space holds for other 
 quasi-probability distributions as well.\\

\noindent
{\em Note Added:} Since completion of this work, a preprint by
Duan et al. \cite{duan} describing an
interesting approach to separability  has appeared. 
 These authors note  that ``the Peres-Horodecki criterion has some
difficulties'' in the continuous case,   
and hence aim at  ``a strong and practical inseparability criterion'',
which  proves   necessary and sufficient in the Gaussian case. 
 We believe that their criterion is unlikely to be any stronger than the
Peres-Horodecki criterion (19). Further, 
  it appears that to apply their criterion  one has to first solve a
pair of nonlinear simultaneous equations to determine  a parameter $a_0$
that enters  their inequality (16). In this sense  the Peres-Horodecki
criterion (19) seems to be easier to implement in  practice; this is  over
and above the merit of   manifest  
invariance under local transformations and mirror reflection it enjoys.

 \noindent
{\bf Acknowledgement}: The author is grateful to  S. Chaturvedi,
R. Jagannathan and N. Mukunda  for insightful comments.

\end{document}